\documentclass[preprint,5p,times,twocolumn]{elsarticle}

\usepackage{graphicx}

\usepackage{amssymb}

  \newcommand{\nn}{\nonumber}

\journal{Physics Letters B}

\begin{document}

\begin{frontmatter}



\title{Top quark effects on the virtual photon structure function at ILC}


\author{Yoshio Kitadono}

\address{Institute of Particle and Nuclear Study, 
 High Energy Accelerator Organization (KEK),\\ 
               1-1 Oho, Tsukuba, Ibaraki, 305-0801, Japan.}

\begin{abstract}
We investigated top quark effects on virtual photon structure functions by pQCD. We include the top quark mass effects on the virtual photon structure function with the quark parton model and with the operator product expansion up to the next-to-leading order in QCD. We also consider the threshold effect on the running coupling constant in the calculation to the effective photon structure function with a matching condition. The numerical calculations are investigated in the kinematical region expected at the future international linear collider. 
\end{abstract}


\end{frontmatter}


\section{Introduction}
The Large Hadron Collider (LHC) has restarted since last year.
One of the most important tasks of the LHC is to discover the 
Higgs particle which will be the origin of the mass of the particles and the other
beyond standard model search is going on now \cite{LHC}. If the new
physics beyond the standard model is discovered, the precise measurement
will be done at the future International Linear Collider (ILC) \cite{ILC}. In
such a case we need to know the background from the standard model,
especially QCD at high energies. 

It is known 
that the two-photon exchange process ($e^{+} + e^{-} \to \gamma^{*} \gamma^{*}
 \to \mbox{hadrons}$)  is dominated over the one-photon
exchange process ($e^{+} + e^{-} \to \gamma^{*} \to
\mbox{hadrons}$) in the electron-positron collision \cite{Twophoton1,Twophoton2}.
The cross section in this two-photon process is characterised by photon
structure functions and see Refs. \cite{Review1,Review2,Review3,Review4} for reviews.  
The photon structure functions have two types,
namely real photon structure functions 
$F^{\gamma}_{2,L}(x,Q^2), g^{\gamma}_{1}(x,Q^2)$ and
virtual photon structure functions 
$F^{\gamma}_{2,L}(x,Q^2, P^2), g^{\gamma}_{1}(x,Q^2,P^2)$, 
where $Q^2=-q^2$ is a squared momentum of the probe photon, 
$P^2=-p^2$ is a squared momentum of the target photon, $x$ is Bjorken variable  
in the two-photon process respectively. While the real photon structure
functions need to include non-perturbative effects like the vector meson
dominance, the perturbative part in the virtual photon
structure functions dominates at the kinematical region 
$\Lambda_{QCD}^2 \ll P^2 \ll Q^2$. Therefore we consider the virtual photon structure functions, especially unpolarised functions, in order to avoid the non-perturbative effect in this Letter. 
  
Much work on the photon structure functions have been carried out for 
both the real photon target \cite{Work_real1,Work_real2,Work_real3,Work_real4,Work_real5,Work_real6,Work_real7,Work_real8,Work_real9} and the virtual photon target \cite{Work_virtual1,Work_virtual2,Work_virtual3,Work_virtual4,Work_virtual5,Work_virtual6}. Although the heavy quark effects on the photon structure functions have been studied \cite{Mass1,Mass2,Mass3,Mass4}, their phenomenological applications were to charm quark or to bottom quark due to the kinematical constraints of experiments. We can expect easily that top quark effects on the photon structure functions will be important at the ILC. 
We know that top quark have $(2e/3)$ charge in the
proton unit of the electro-magnetic charge and the top quark is the
heaviest quark in the standard model. 
The large electro-magnetic
charge of up-type quarks compared to down-type quarks relatively will enhance the value of the photon structure function, 
but the large mass of the top quark will reduce that of
photon structure functions.  We have to study the size of the top quark
effects on photon structure functions at the ILC.

In this Letter, we consider the top quark effects
on the unpolarised virtual photon structure functions with the method based on the operator product expansion (OPE) improved by the renormalisation group equation (RGE), and with the method based on the quark parton model (QPM). The top quark effects by OPE and QPM at ILC are discussed in the next section and the numerical calculation based on the framework with OPE
and QPM are discussed and the results are shown in the section \ref{results}. 
Final section is devoted to a conclusion. 

\section{Top quark effects on virtual photon structure functions} \label{top}
We can incorporate the top quark effects by the two methods, 
namely OPE supplemented by the RGE and QPM. 
We use the formalism in Ref. \cite{HQEOPE} for the calculation of photon
structure functions by OPE and we use the results in Ref. \cite{Mass1} by QPM.

\subsection{Operator Product Expansion}
Let us define the $n$-th moment of photon structure functions by the equation
\begin{eqnarray}
 M^{\gamma}_{2(L)}(n,Q^2,P^2) 
  &=& \int_{0}^{1} dx x^{n-2} F^{\gamma}_{2(L)}(x,Q^2,P^2),
\end{eqnarray}
where $n-2$ is due to the our convention of structure functions.
In the formalism of Ref. \cite{HQEOPE}, it is assumed that we divide the $n_f$
quarks system into two parts, namely $n_f-1$ massless quarks system and
one massive quark. We apply this formalism to the virtual photon
structure functions at the ILC and we assume that $u,d,s,c,b$ are massless
quarks and $t$ is the massive quark at the kinematical region expected at the ILC.
The moment by OPE including heavy quark effects can be summarised as the
following form,
\begin{eqnarray}
&{}& \hspace{-1.0cm} M^{\gamma}_{2(L)}(n,Q^2,P^2) \nn\\
 &=&   M^{\gamma}_{2(L)}(n,Q^2,P^2) {}_{\mbox{\tiny massless}}
    + \Delta M^{\gamma}_{2(L)}(n,Q^2,P^2),
\end{eqnarray}
where the first term means the contribution from massless quarks and is
given as
\begin{eqnarray}
&{}& \hspace{-1cm}  M^{\gamma}_{2(L)}(n,Q^2,P^2) {}_{\mbox{\tiny massless}} 
   / \frac{\alpha}{8\pi\beta_0} \nn\\
 &=& \frac{4\pi}{\alpha_s(Q^2)} 
       \sum_{i=\pm,NS} \mathcal{L}_{2(L),i}^{n}
       \left( 1 - r^{d_{i}^{n} + 1} \right)
\nn\\ 
&{}& 
     + \sum_{i=\pm,NS} \mathcal{A}_{2(L),i}^{n}
       \left( 1 - r^{ d_{i}^{n} } \right)
     + \sum_{i=\pm,NS} \mathcal{B}_{2(L),i}^{n}
       \left( 1 - r^{d_{i}^{n} + 1} \right) 
\nn\\ 
&{}& 
     + \mathcal{C}_{2(L)}^{n}, \label{mom_massless}
\end{eqnarray}
where $r=\alpha_s(Q^2)/\alpha_s(P^2)$ is the ratio of coupling constant with different scales, $~d_{i}^{n}$ corresponds to the eigen-values of one-loop hadronic anomalous dimension matrix, 
the sum runs over the index to the same eigen-values.
These forms of the perturbative expansion are common for $M^{\gamma}_{2}$ and
$M^{\gamma}_{L}$ up to the next-to-leading order (NLO) in QCD. 
The long expression to the coefficients
$\mathcal{L}^{n}_{2(L),i}, \mathcal{A}^{n}_{2(L),i},
\mathcal{B}^{n}_{2(L),i}$ and $\mathcal{C}_{2(L)}^{n}$ are given in Ref. \cite{Work_virtual1}.
The above moment consists of $n_f-1$ massless quarks ($u,d,s,c,b$).

On the other hand, the heavy quark effects are incorporated in the
moment of the effective structure function up to the NLO in QCD with OPE 
supplemented by the mass-independent RGE formalism and the moment is given by the form,
\begin{eqnarray}
&{}& \hspace{-1cm}  \Delta M^{\gamma}_{2}(n,Q^2,P^2)
   / \frac{\alpha}{8\pi\beta_0} \nn\\
  &=& \sum_{i=\pm,NS} \Delta \mathcal{A}_{i}^{n}
       \left( 1 - r^{ d_{i}^{n} } \right)
     + \sum_{i=\pm,NS} \Delta \mathcal{B}_{i}^{n}
       \left( 1 - r^{d_{i}^{n} + 1} \right)
\nn\\ 
&{}& 
     + \Delta \mathcal{C}^{n},
\end{eqnarray}
where the expression of the coefficients 
$\Delta \mathcal{A}^{n}_{i}, \Delta \mathcal{B}^{n}_{i}$ and $\Delta
\mathcal{C}^{n}$ are given in Ref. \cite{HQEOPE} and all finite
coefficients are related with the variation of the operator matrix
element for the top quark due to mass effects.
The variation by the heavy quark effects to coefficients in 
$M_{L}^{\gamma}$ is zero up to this order. 
The above variation of the moment due to the top quark mass
consists of one-massive quark $(t)$ as we mentioned previously.
We reconstruct the structure functions from the moment by
Mellin inversion numerically,
\begin{eqnarray}
F^{\gamma}_{2(L)}(x,Q^2,P^2) 
  &=& \frac{1}{2\pi i}
      \int_{c-i\infty}^{c+i\infty} 
      \hspace{-0.5cm} dn x^{-n-1} 
      M^{\gamma}_{2(L)}(n,Q^2,P^2), 
\end{eqnarray} 
where $c$ is a positive constant. Although we choose $c=1.5$, generally speaking, 
the result is independent of choice of constant $c$.

\subsection{Quark Parton Model}
The effects of the heavy quark mass are incorporated by the heavy quark propagator in related QED box diagrams. The structure functions by QPM are given by the equations,
\begin{eqnarray}
 F_{2}^{\gamma}\Big|^{\mbox{\tiny QPM}} 
  &=& \frac{x}{\tilde{\beta^2}}
      \left(W_{TT} + W_{LT} 
                   - \frac{1}{2} W_{TL} 
                   - \frac{1}{2} W_{LL} \right),\\
 F_{L}^{\gamma}\Big|^{\mbox{\tiny QPM}} 
  &=& x\left(W_{TT} - \frac{1}{2} W_{LL}\right),
\end{eqnarray}
where $\tilde{\beta} = \sqrt{1-p^2q^2/(p\cdot q)^2}$ and the explicit
expressions of $W_{TT},W_{LT},W_{TL},$ and $W_{LL}$ are given by the
equations of Appendix B in Ref. \cite{Mass1}. 
Although the above normalisation is different from one in Ref. \cite{Mass1}, 
we use $2W_{TT},...,2W_{LL}$ in Ref. \cite{Mass1}. This convention is compatible with the normalisation used in Ref. \cite{HQEOPE}.
In QPM results, all structure functions $W_{TT}, ..., W_{LL}$ are
expressed by the factors $\beta$, $\tilde{\beta}$, $L$, $Q^2, P^2$ and $x$.
The parameters $\beta$ and $L$ are given by
$ \beta = \sqrt{1 - \frac{(4m^2+P^2)}{Q^2}\frac{x}{(1-x)} }, $
$L = \log \left(\frac{1+\beta}{1-\beta} \right),$
and the threshold effect of the heavy quark mass is controlled by these factors.
The factors $\beta$ and $L$ vanish at a maximum point of Bjorken variable
$ x_{\max}=\frac{1}{1+\frac{4m^2+P^2}{Q^2}}, $
where this maximum point in Bjorken $x$ is derived by the condition 
$s = (p + q)^2 \ge 4m^2$.
Therefore structure functions $F^{\gamma}_{2}, F^{\gamma}_{L}$ by QPM
are limited in the range $0\le x \le x_{\max}$ and are insured to vanish
at the points $x=0,~x_{\max}$, but the structure functions by OPE is not
guaranteed to vanish at the points. 

\section{Numerical Calculations \label{results}}
The effective photon structure function is often measured in
experiments. This effective structure function is proportional to the
total cross section of the two-photon process and is given by the equation,
\begin{eqnarray}
F^{\gamma}_{\mbox{\tiny eff}}(x, Q^2,P^2) 
= F^{\gamma}_{2}(x, Q^2,P^2)
+ \frac{3}{2} F^{\gamma}_{L}(x, Q^2,P^2).
\end{eqnarray}
We used following masses for both QPM and OPE as inputs in this Letter,
\begin{eqnarray}
 m_{u} &=& 0.003 ~\mbox{\rm GeV}, \hspace{1cm}
 m_{d}  = 0.006  ~\mbox{\rm GeV},\nn\\
 m_{s} &=& 0.12  ~\mbox{\rm GeV},\hspace{1.2cm}
 m_{c}  =  1.3   ~\mbox{\rm GeV},\nn\\
 m_{b} &=& 4.2   ~\mbox{\rm GeV},\hspace{1.4cm}
 m_{t}  = 170    ~\mbox{\rm GeV}, \label{inputs}
\end{eqnarray}
where we consider all quarks are massive in QPM and the top quark is the
massive particle in OPE.

\subsection{$Q^2$ and $P^2$ dependence in a moment of the effective photon structure function}
\begin{figure}
	\begin{center}
	\def\SCALE{0.65}
	\includegraphics[scale=\SCALE]{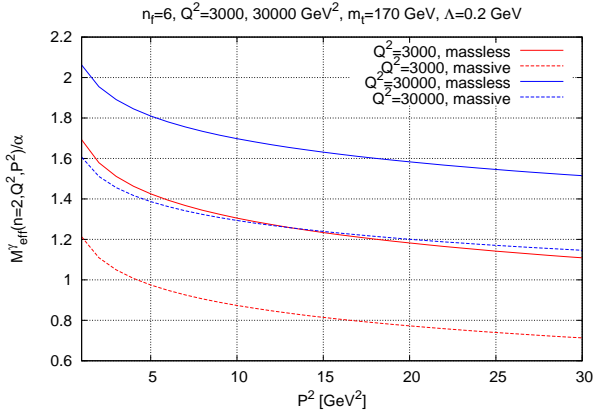} 
		\caption{ $P^2$ dependence in the moment of the effective structure function for  
   $Q^2=3000~$GeV$^2$, $Q^2=30000~$GeV$^2$. 
	The lines labelled 'massless (massive)' are the results without (with) the top quark mass effects.
    }
    \label{fig1}
	\end{center}	
\end{figure}

We calculate the $P^2$ dependence to the moment with $n=2,
Q^2=3000~\mbox{GeV}^2, Q^2=30000~\mbox{GeV}^2$ by OPE ,
\begin{eqnarray}
 M^{\gamma}_{\mbox{\tiny eff}}(n=2, Q^2,P^2) 
= \int_{0}^{1}dx F^{\gamma}_{\mbox{\tiny eff}}(x, Q^2,P^2),
\end{eqnarray}
and it is shown in Fig.\ref{fig1}.
The value of the moment decrease as $P^2$ (target virtuality) increase slowly.
Because the target virtuality corresponds to the mass of the target photon, 
the phase-space suppression due to the target mass reduces the moment.
The dominant contribution to $P^2$ dependence of the moment comes from
the term $r \approx 1/\ln(P^2/\Lambda^2_{\mbox{\tiny QCD}})$ in
the eq.(\ref{mom_massless}). The $P^2$ dependence in the heavy quark
effects by OPE appears in the coefficients $\mathcal{A}, \mathcal{C}$ 
to the moment in eq.(\ref{mom_massless})
through the heavy quark operator matrix element and it behaves as
$\ln(P^2/m^2)$. 
We compare the results with and without heavy
quark effects by OPE. We can see the tendency that the $P^2$ dependence
with $Q^2=30000~\mbox{GeV}^2$ including the top quark effects is similar
to the result with $Q^2=3000~\mbox{GeV}^2$ neglecting the top quark
effects. 
Comparing the massive results with the massless results, 
the size of top quark mass effects on the $P^2$ dependence is
about $40~\%$ to $55~\%$ for the case $Q^2=3000~\mbox{GeV}^2$ and about
$30~\%$ for the case $Q^2=30000~\mbox{GeV}^2$. 
 
\begin{figure}
	\begin{center}
	  \def\SCALE{0.65}	
      \includegraphics[scale=\SCALE]{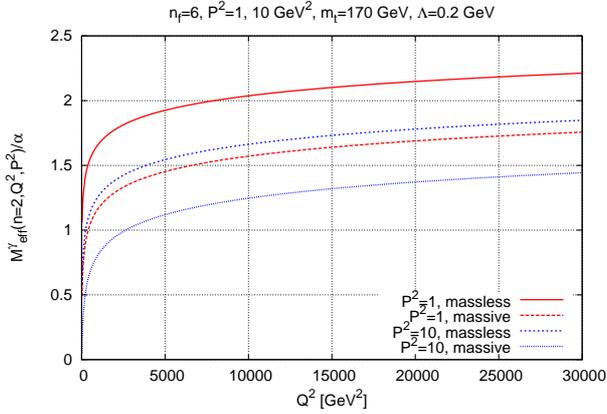}
    \caption{ $Q^2$ dependence in the moment of the effective structure function for  
   $P^2=1~$GeV$^2$, $P^2=10~$GeV$^2$. 
	The lines labelled 'massless (massive)' are the results without (with) the top quark mass effects.   
    }
	\label{fig2}
    \end{center}
\end{figure}

On the other hand, 
the $Q^2$ dependence of $M^{\gamma}_{\mbox{\tiny eff}}(n=2, Q^2,P^2)$
with $P^2=1, 10~\mbox{GeV}^2$ is shown in Fig.\ref{fig2}.
The value of the moment increase as $Q^2$ (resolution) increase, since
the higher $Q^2$ corresponds to the better resolution in the two-photon
process and it emerges many partons in the virtual photon. 
The dominant contribution to the $Q^2$ dependence comes from
the term $4\pi/\alpha_s(Q^2) \approx \ln(Q^2/\Lambda^2_{\mbox{\tiny QCD}})$
in the eq.(\ref{mom_massless}).
Comparing the massive results with the massless results, 
the size of top quark mass effect on the $Q^2$ dependence falls below
$30~\%$ in the region $Q^2 > 1000~\mbox{GeV}^2$ for the case
$P^2=1~\mbox{GeV}^2$ and falls below $30~\%$ in the region $Q^2 > 3000
\mbox{GeV}^2$ for the case $P^2=10~\mbox{GeV}^2$.

\subsection{Threshold effects on the effective photon structure function}
\begin{figure}
  \begin{center}
    \def\SCALE{0.65}
      \includegraphics[scale=\SCALE]{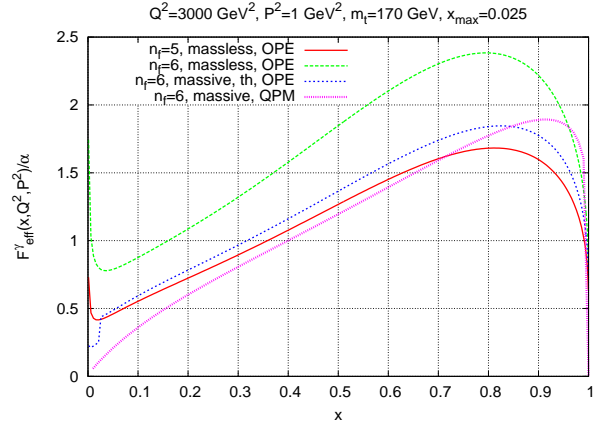} 
   \caption{ 
Effective structure function with $Q^2=3000$~GeV$^2$, $P^2=1$~GeV$^2$. 
The lines labelled '$n_f=6$, massive, th, OPE' and '$n_f=6,$ massive, QPM' are the results by  OPE with threshold effects and the results by QPM respectively. The other lines are the massless results by OPE with $n_f=5,6$.
    }
\label{fig3}
  \end{center}
\end{figure}

\begin{figure}
  \begin{center}
    \def\SCALE{0.65}
      \includegraphics[scale=\SCALE]{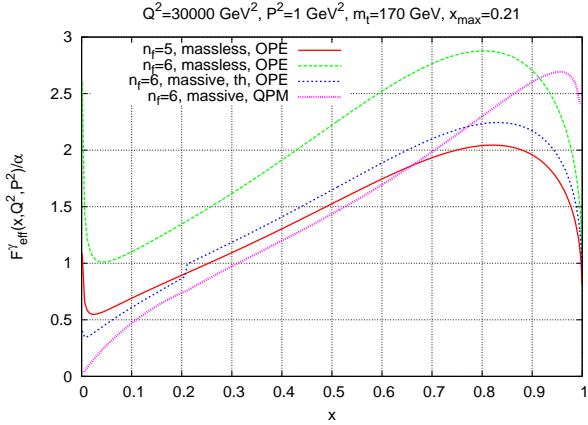} 
   \caption{ 
Effective structure function with $Q^2=30000$~GeV$^2$, $P^2=1$~GeV$^2$. 
The lines labelled '$n_f=6$, massive, th, OPE' and '$n_f=6,$ massive, QPM' are the results by  OPE with threshold effects and the results by QPM respectively. The other lines are the massless results by OPE with $n_f=5,6$.
    }
\label{fig4}
  \end{center}
\end{figure}

The threshold effect due to the top quark mass is included by a 
matching condition to the running-coupling constant 
(for example see Refs. \cite{HQE.alphas1, HQE.alphas2} )
\begin{eqnarray}
 \alpha_s^{n_f=5}(m_t^2) 
= \alpha_s^{n_f=6}(m_t^2)( 1 + O(\alpha^2_s)).
\end{eqnarray}
Equivalently we convert the above equation into the relation between 
$\Lambda_{n_f=5}$ and $\Lambda_{n_f=6}$ up to the NLO in QCD, 
we find that $\Lambda_{n_f=6}=0.080 ~\mbox{GeV}$ 
for $\Lambda_{n_f=5}=0.20 ~\mbox{GeV}$. 
We used this matching condition in the calculation by OPE. In addition
to this condition, we change the number of active flavors in the moment 
at the threshold point $x=x_{\max}$.
We evaluate the effective structure function including
the top quark mass effects with $n_f=6$ theory in the region $x<x_{\max}$,
and evaluate the structure function without the top quark
effect with $n_f=5$ theory in the region $x_{\max}\le x$.
Our numerical calculations to the effective photon structure function
are shown in Fig.\ref{fig3} for $Q^2 = 3000~\mbox{GeV}^2$,  
and in Fig.\ref{fig4} for $Q^2 = 30000~\mbox{GeV}^2$.

Although the result with massless $n_f=6$ quarks are not reduced by the phase-space suppression due to the heavy quark effects, results with massless
$n_f=5$ quarks and one-massive quark (top) are reduced by the suppression in 
Fig.\ref{fig3} and in Fig.\ref{fig4}. We also plot the
result by $n_f=5$ massless quarks by OPE as reference.  
We can see that the results by OPE and the results by QPM are similar
and consistent each other. 

While the threshold effect at $x_{\max}=0.025,~0.21$ cannot be seen in the results (the line labelled '$n_f=6,$massive QPM' in Fig.\ref{fig3} and Fig.\ref{fig4}), we find a jump corresponding to the top quark threshold at
$x_{\max}$ in the results ('$n_f=6,$ massive,th,OPE' in Fig.\ref{fig3} and in Fig.\ref{fig4}). We can see that the result by OPE and that by QPM are similar results each other.

Comparing the massive result ('$n_f=6,$ massive,th,OPE' in Fig.\ref{fig3} and Fig.\ref{fig4}) with the massless result ('$n_f=6,$ massless,OPE' in Fig.\ref{fig3} and Fig.\ref{fig4}), 
the size of top quark mass effects on the effective virtual photon
structure function is about $70~\%$ to $90~\%$ in the region $x<x_{\max}$
for the case $Q^2=30000~\mbox{GeV}^2$ and about $35~\%$ to $85~\%$ in the region
$x<x_{\max}$ for the case $Q^2=3000~\mbox{GeV}^2$. 
But we can say that the top quark almost decouples for the case $Q^2=3000~\mbox{GeV}^2$ due to the smallness of the range $x<x_{\max}$.

\section{Conclusion} \label{conclusion}
We have investigated the top quark effects on the virtual photon structure
functions by OPE and by QPM with the kinematical region expected at the future ILC. 
The top quark effects are incorporated in the matrix element of moments by OPE formalism and are included in the Feynman diagram by QPM.  
We evaluated $P^2$ and $Q^2$ dependence to a fixed moment of the effective virtual photon structure function. Our calculation shows that the $P^2$ dependence including top quark mass effects at $P^2= 1~\mbox{GeV}^2, Q^2=30000~\mbox{GeV}^2$ are similar to the results without top quark mass effects at $P^2= 1~\mbox{GeV}^2, Q^2=3000~\mbox{GeV}^2$. The situation to $Q^2$ dependence is similar.

We also evaluated the theoretical calculation to the effective virtual photon structure function.  In this calculation we added the threshold effect in the NLO running-coupling constant with the matching condition. The top quark mass effects are rather large at the kinematical region expected at ILC.
We also find the consistency between the calculations by OPE and that by QPM. 

If the effective photon structure function or the total cross section in two-photon process by double-tagging electron and positron are measured at the ILC, we can compare the theoretical calculations with the experimental data. Then we will discuss the validity of QCD by $e^{+}+e^{-}$ collider around $\mbox{TeV}$ scale with photon structure functions and we might study the physics about top quark and related topics in two-photon process. Furthermore the study on the polarised photon structure functions at ILC will be interesting topic by using the planned polarisation option. 

\section*{Acknowledgements}
I would like to thank Tsuneo Uematsu (Kyoto U.) for reading the manuscript. 
I also thank Ken Sasaki (Yokohama National U.), Hiroyuki Kawamura (KEK) and Takahiro Ueda (KIT) for useful discussion. 
%
%








\end{document}